\documentclass[prl,showpacs,twocolumn]{revtex4}
\usepackage{graphics}
\usepackage{epsfig}
\usepackage{graphicx}

\begin{document}
\title{General Sequential Quantum Cloning}
\author{Gui-Fang Dang and Heng Fan
 }
\affiliation{Institute of Physics,
Chinese Academy of Sciences, Beijing 100080, China.
}

\date{\today}

\begin{abstract}
Some multipartite quantum states can be generated in a sequential manner which may be implemented
by various physical setups like microwave and optical cavity QED, trapped ions, and quantum dots etc. 
We analyze the general $N$ to $M$ ($N\le M$) qubits Universal Quantum Cloning Machine (UQCM) 
within a sequential
generation scheme. We show that the $N$ to $M$ sequential UQCM is available. 
The case of $d$-level quantum states sequential cloning is also presented.  
\end{abstract}

\maketitle

Quantum entanglement plays a key role in quantum computation and quantum information \cite{BD}.
Multipartite entangled states arise as a resource for quantum information processing tasks
such as the well known quantum teleportation\cite{BBCJPW}, quantum communication \cite{GC,RB},
clock synchronization \cite{GLM} etc.
In general it is extremely difficult to generate experimentally multipartite entangled states through
single global unitary operations. In this sense, the {\it sequential} generation of the entangled 
states appears to be promising. Actually most of the quantum computation networks are designed
to implement quantum logic gates through a sequential procedure \cite{BBCD}. 
Recently sequential implementing of quantum information
processing tasks has been attracting much attention. It is pointed out that photonic
multiqubit states can be generated by letting a source emit photonic qubits in a sequential
manner \cite{SGTCZ}. The general sequential generation of entangled multiqubit states in the
realm of cavity QED was systematically studied in Refs.\cite{SSVCW,SHWCS}. 
It is also shown that the class of sequentially
generated states is identical to the matrix-product-state (MPS) which is very useful in study of
spin chains of condensed matter physics \cite{AKLT}.

On the other hand, much progress has already been made in the past years in studying quantum 
cloning machines, for reviews see, for example, Refs.\cite{SIGA,CF,Fan}. And various
quantum cloning machines have been implemented experimently by polarization of photons 
\cite{LSHB,MBSS,PSSSD,ILDB,RSSD},nuclear spins in Nuclear Magnetic Resonance \cite{CJFS,Du},
etc. However, these experiments are for 1 to 2 (one qubit input and two-qubit output) or 1 to 3
cloning machines. The more general case will be much difficult. 
There are some schemes proposed for the
general quantum cloning machines which are not in a sequential manner, see for example,
\cite{SWZ,FWMW}. Recently a 1 to $M$ sequential universal quantum cloning 
is proposed \cite{DLLSS} by using the cloning transformation presented in Ref.\cite{GM}.
Since it is in a sequential procedure, 
potentially it reduces the difficult in implementing this quantum cloning machine.
However,
as is well known the collective quantum cloning machine (the $N$ identical input states
are cloned collectively to $M$ copies) is better than the quantum cloning machine which
can only deal with the individual input(only one input is copied to several copies each time). 
We know that the general $N$ to $M$ cloning transformation is also available 
in Refs.\cite{GM,FMW}. Then a natural question arise is that whether the general $N$ to
$M$ {\it sequential} cloning machine is possible. 
In this Letter, we will present the general sequential universal quantum cloning machine. 

The 1 to $M$ cloning transformations used in Ref.\cite{DLLSS} 
was proposed by Gisin and Massar in Ref.\cite{GM}. And the 
$N$ to $M$ UQCM was also presented in Ref.\cite{GM}. However, to use the 
method proposed in Refs.\cite{DLLSS,SSVCW} to find the sequential cloning machine, the input
state $|\Phi \rangle ^{\otimes N}$ 
should be expanded in computational basis $\{|0\rangle ,|1\rangle \}$.
The explicit quantum cloning transformations with this kind of input were 
proposed by Fan {\it et al}
in Ref.\cite{FMW}. In this Letter, based on the result of Ref.\cite{FMW}, the general
sequential UQCM will be presented.

As presented in Refs.\cite{SSVCW,DLLSS}, the sequential generation of a multiqubit state is like the
following. Let ${\cal {H_A}}$ be a $D$-dimensional Hilbert space which acts as the ancillary system,
and a single qubit (e.g., a time-bin qubit) is in a two-dimensional Hilbert space ${\cal {H_B}}$. 
In every step of the sequential generation of a multiqubit state, a unitary time evolution
will be acting on the joint system ${\cal {H_A}}\otimes {\cal {H_B}}$. We assume that
each qubit is initially in the state $|0 \rangle $ which is like a blank or an empty state 
and will not be written out in the formulas.
So the unitary time evolution is written in the form of an isometry $V$: 
${\cal {H_A}}\rightarrow {\cal {H_A}}\otimes {\cal {H_B}}$, where 
$V=\sum _{i, \alpha ,\beta }V_{\alpha ,\beta }^i|\alpha ,i\rangle \langle \beta | $,
each $V^i$ is a $D\times D$ matrix, and the isometry condition takes the
form $\sum _{i=0}^1V^{i\dagger }V^i=1$. By applying successively $n$ operations of $V$
(not necessarily the same) 
on an initial ancillary state $|\phi _I\rangle \in {\cal {H_{A}}}$, we obtain
$|\Psi \rangle =V^{[n]}...V^{[2]}V^{[1]}|\phi _I\rangle $. The generated $n$ qubits are 
in general an entangled state, but the last step qubit-ancilla interaction can be chosen
so as to decouple the final multiqubit entangled state from the auxiliary system, 
so the sequentially generated state is 
\begin{eqnarray}
|\psi \rangle =\sum _{i_1...i_n=0}^1\langle \phi _F|
V^{[n]i_n}...V^{[1]i_1}|\phi _I\rangle |i_n,...,i_1\rangle ,
\label{MPS}
\end{eqnarray}  
where $|\phi _F\rangle $ is the final state of the ancilla. 
This is the MPS. It was proven that any MPS can be sequentially generated \cite{SSVCW}. 

Suppose there are $N$ identical pure quantum states $|\Phi \rangle ^{\otimes N}=
(x_0|0\rangle +x_1|1\rangle )^{\otimes N}$
need to be cloned to $M$ copies, where $|x_0|^2+|x_1|^2=1$. We know that the input state can
be represented by a basis in symmetric subspace.
\begin{eqnarray}
|\Phi \rangle ^{\otimes N}=\sum _{m=0}^Nx_0^{N-m}x_1^m
\sqrt {C_N^m}|(N-m)0, m1\rangle ,
\end{eqnarray}
where $|(N-m)0,m1\rangle $ denotes the symmetric and normalized state with $(N-m)$ qubits
in the state $|0\rangle $ and $m$ qubits in the state $|1\rangle $, and we have
$C_N^m=N!/(N-m)!m!$ in standard notation. So if we find the quantum cloning transformations
for all states in symmetric subspace, we can clone $N$ pure states to $M$ copies. 
The UQCM with input in symmetric subspace
can be written as \cite{FMW},
\begin{eqnarray}
|(N-m)0, m1\rangle &\rightarrow &|\Phi ^m_M\rangle ,
\end{eqnarray}
where
\begin{eqnarray}
|\Phi ^m_M\rangle &=& \sum _{j=0}^{M-N}\beta _{mj}|(M-m-j)0,(m+j)1\rangle 
\otimes R_j,
\label{clone}
\\
\beta _{mj}&=&\sqrt {C_{M-m-j}^{M-N-j}C_{(m+j)}^j/C_{M+1}^{N+1}},
\label{clone1}
\end{eqnarray}
where $R_j$ are the ancillary states of the cloning machine
and are orthogonal with each other for different $j$.
For a sequential quantum cloning machine in this Letter,
we choose a realization $R_j\equiv |(M-N-j)1,j0\rangle $ for the ancilla states. 
This UQCM is optimal in the sense that the fidelity between single qubit output state
reduced density operator $\rho ^{out}_{reduced}$ and the single input $|\Phi \rangle $
is optimal. The optimal fidelity is $F=\langle \Phi |\rho _{reduced}^{out}|\Phi \rangle 
=(MN+M+N)/M(N+2)$, see Refs.\cite{SIGA,CF,Fan} for reviews and the references therein.
A realization of this UQCM with photon stimulated emission can be found in 
Ref.\cite{FWMW} which is not in a sequential manner. We next show that this general
$N$ to $M$ UQCM can be generated through a sequential procedure. 

The basic idea is to show that the final state of the cloning, 
$|\Phi ^m_M\rangle $ in (\ref{clone}),
can be expressed in its
MPS form. As shown in Ref.\cite{SSVCW}, any MPS can be sequentially generated.
We shall follow the method, for example, as in Refs.\cite{Vidal,DLLSS}.
By Schmidt decomposition, we first express the quantum state
$|\Phi _M^m\rangle $ as a bi-partite state across $1:2...$ cut,
\begin{eqnarray}
|\Phi _M^m\rangle &=&\lambda _1^{[1]}|0\rangle |\phi _1^{[2...(2M-N)]}\rangle 
+\lambda _2^{[1]}|1\rangle |\phi _2^{[2...(2M-N)]}\rangle 
\nonumber \\
&=&\sum _{\alpha _1i_1}\Gamma ^{[1]i_1}_{\alpha _1}\lambda ^{[1]}_{\alpha _1}|i_1\rangle 
|\phi _{\alpha _1}^{[2...(2M-N)]}\rangle ,
\end{eqnarray}
where $\Gamma ^{[1]0}_{\alpha _1}=\delta _{\alpha _1,1}, 
\Gamma ^{[1]1}_{\alpha _1}=\delta _{\alpha _1,2}$, and 
$\lambda ^{[1]}_{\alpha _1}$ are eigenvalues of the first qubit reduced density
operator, and we find
$\lambda _1^{[1]}=\sqrt {\sum _{k=-m}^{M-m-1}\beta _{mk}^2C_{M-1}^{m+k}/C_M^{m+k}}$,
$\lambda _2^{[1]}=\sqrt {\sum _{k=-m}^{M-m-1}\beta _{mk+1}^2C_{M-1}^{m+k}/C_M^{m+k+1}}$.
To correspond with the MPS in (\ref{MPS}), we can define
$V_{\alpha _1}^{[1]i_1}=\Gamma _{\alpha _1}^{[1]i_1}\lambda ^{[1]}_{\alpha _1}$.
Successively by Schmidt decomposition, the quantum state $|\Phi _M^m\rangle $ in 
(\ref{clone}) is divided into a
bi-partite state with the first $n$ qubits as one part, and the rest as
another part, where $1<n\le M-1$. We find
\begin{eqnarray}
|\Phi _M^m\rangle =\sum _{j=0}^{n'}\lambda ^{[n]}_{j+1}|(n-j)0,j1\rangle 
|\phi _{j+1}^{[(n+1)...(2M-N)]}\rangle ,
\label{schmidt}
\end{eqnarray} 
when $1<n\le M-N+m, n'=n$; when $M-N+m<n
\le M-1, n'=M-N+m$, $\lambda ^{[n]}_{j+1}$ are eigenvalues of the
first $n$ qubits reduced density operator of $|\Phi ^m_M\rangle $. 
According to the results in 
Eqs.(\ref{clone},\ref{clone1}), we can obtain, 
\begin{eqnarray}
\lambda ^{[n]}_{j+1}=\sqrt {C_n^j\sum _{k=-m}^{M-m-n}\beta _{m(j+k)}^2
\frac {C_{M-n}^{m+k}}{C_M^{m+j+k}}}.
\end{eqnarray}
And we also have 
\begin{eqnarray}
&&|\phi _{j+1}^{[(n+1)...(2M-N)]}\rangle 
=\frac {\sqrt {C_n^j}}{\lambda ^{[n]}_{j+1}}\sum _{k=-m}^{M-m-n}
\beta ^2_{m(j+k)}\times
\nonumber \\
&&\times \sqrt {\frac {C_{M-n}^{(m+k)}}{C_M^{m+j+k}}}
|(M-n-m-k)0,(m+k)1\rangle \otimes R_{j+k} .
\nonumber
\end{eqnarray}
By induction and a concise formula, we have 
\begin{eqnarray}
&&|\Phi _{j+1}^{n...(2M-N)]}\rangle 
\nonumber \\
&=&
\sum _{\alpha _n,i_n}\Gamma ^{[n]i_n}_{(j+1)\alpha _n}\lambda ^{[n]}_{\alpha _n}
|i_n\rangle |\phi _{\alpha _n}^{[(n+1)...(2M-N)]}\rangle  
\nonumber \\
&=&\frac {\sqrt {C_{n-1}^j}}{\lambda ^{[n-1]}_{j+1}}\left[
|0\rangle |\phi _{j+1}^{[(n+1)...(2M-N)]}\rangle 
\frac {\lambda ^{[n]}_{j+1}}{\sqrt {C_n^j}}
\right.
\nonumber \\
&&\left. +|1\rangle |\phi _{j+2}^{[(n+1)...(2M-N)]}\rangle 
\frac {\lambda ^{[n]}_{j+2}}{\sqrt C_n^{j+1}}\right] ,
\end{eqnarray}
where we denote 
\begin{eqnarray}
\Gamma ^{[n]0}_{(j+1)\alpha _n}&=&\delta _{(j+1)\alpha _n}
\sqrt {C_{n-1}^j}/(\lambda ^{[n-1]}_{j+1}\sqrt {C_n^j}),
\\
\Gamma ^{[n]1}_{(j+1)\alpha _n}&=&\delta _{(j+2)\alpha _n}
\sqrt {C_{n-1}^j}/(\lambda ^{[n-1]}_{j+1}\sqrt {C_n^{j+1}}).
\end{eqnarray}
Still we define that
\begin{eqnarray}
V^{[n]i_n}_{\alpha _n\alpha _{n-1}}=\Gamma ^{[n]i_n}_{\alpha _{n-1}\alpha _n}
\lambda ^{[n]}_{\alpha _n}.
\end{eqnarray}
It is thus in the MPS representation. We can further consider other
cases including the ancilla state of the cloning machine
represented as $R_j$ (Note it is not the ancilla state in the MPS
representation).  We can find that the output state of the general UQCM can be
expressed as MPS as in form (\ref{MPS}). So it can be
created sequentially. The explicit results are summarized in the appendix.

We have shown that the output states of the general UQCM in (\ref{clone},\ref{clone1})
are MPS's and thus can be generated sequentially. The sequential matrices $V^{[n]}$
of course depend on the input $|(N-m)0,m1\rangle $ which are W-like states
and are generally multiqubit entangled. For later convenience, we denote $V(m)$ to
express that it depends on input state for different $m$.
By a straightforward method, the sequential cloning operation, 
i.e., the isometrices, depending on
different input may take the form 
$\sum _m|(N-m)0,m\rangle \langle (N-m)0,m1|\otimes V(m)$. However, this
operation may need a single global unitary operator which involves $N$-qubit 
entangled states except for $m=0,m=N$. This contradicts with our aim
that each operation should be divided into sequential unitary operators in a 
{\it quDit} (quantum state in $D$-dimensional space) times qubit system.  
Here we can use a scheme like the following: the ancillary state interacts with
each qubit according to the $(N+1)\times D$-dimensional isometrices as
$\sum _m\sqrt {C_N^m}|0\rangle \langle 0|^{\otimes N-m}\otimes 
|1\rangle \langle 1|^{\otimes m}\otimes V(m)$ sequentially, here
a whole normalization factor is omitted. 
We know that the operation $|0\rangle \langle 0|^{\otimes N-m}\otimes 
|1\rangle \langle 1|^{\otimes m}$ acts on each qubit individually. 
Thus this scheme reduces the complexity of the operation.
This finishes our general sequential UQCM for the case of qubit.
In case $N=1$, we recover the result of Ref.\cite{DLLSS} for 1 to $M$
cloning. We should remark that similar as the
case of sequential 1 to $M$ UQCM in Ref.\cite{DLLSS},
for the general sequential UQCM, the minimal dimension $D$ of the ancillary state grows
linearly at most with $M-N/2+1$ for even $N$ or $M-(N-1)/2$ for odd $N$.

Next we will consider a more general case that 
the sequential cloning machine
is about the quantum state
in $d$-dimensional Hilbert space. 
We will use the
$d$-dimensional UQCM proposed by Fan {\it et al}
in Ref. \cite{FMW}. This UQCM is a generalization of the
cloning machine proposed in Ref.\cite{GM} and
we can use this UQCM to study its sequential form 
for $d$-dimensional case.     

An arbitrary $d$-dimensional pure state takes the form 
$|\Phi \rangle =\sum _{i=0}^{d-1}x_{i}|i\rangle $ with $\sum _{i=0}^{d-1}|x_i|^2=1$.
$N$ identical pure states can be expanded in terms of state in symmetric subspace
$|\Phi \rangle ^{\otimes N}=\sum _{\vec {m}=0}^N
\sqrt {\frac {N!}{m_1!...m_d!}}x_0^{m_1}...x_{d-1}^{m_d}|\vec {m}\rangle $,
where $|\vec {m}\rangle \equiv |m_1,...,m_d\rangle $ is a symmetric state with
$m_i$ states of $|i-1\rangle $, and also $m_i$ should satisfy a relation
$\sum _{i=1}^dm_i=N$.   
The cloning transformations with states in symmetric subspace can be written as
\begin{eqnarray}
|\vec {m}\rangle &\rightarrow &|\Phi _M^{\vec {m}}\rangle =\sum _{\vec{j}=0}^{M-N}
\beta _{\vec {m}\vec {j}}|\vec{m}+\vec{j}\rangle \otimes |\vec{j}\rangle ,
\\
\beta _{\vec {m}\vec {j}}&=&
\sqrt {\frac {\prod _{i=1}^dC_{m_i+j_i}^{j_i}}{C_{M+d-1}^{M-N}}}
\end{eqnarray}
where $\vec {j}$ should satisfy $\sum _ij_i=M-N$.
This cloning machine is optimal and the corresponding fidelity of a single
quantum state between input and output is 
$F=\left( N(d+M)+M-N\right) /(d+N)M$.

As for qubit system, we next show that the output states for all symmetric
states input can be expressed as the sequential form.
We consider the case $1<n\le M-1$, and the state $|\Phi _M^{\vec {m}}\rangle $
is a bipartite state across $1...n:(n+1)... $ cut,
\begin{eqnarray}
|\Phi _M^{\vec {m}}\rangle =\sum _{\vec {j}=0}^M\sum _{\vec {k}=0}^{M-n}
\lambda ^{[n]}_{\vec {j}}|\vec {j}\rangle |\phi _{\vec {j}}^{[(n+1)...(M+1)]}\rangle 
\end{eqnarray}
where
\begin{eqnarray}
\lambda ^{[n]}_{\vec {j}}=\sqrt 
{\sum _{\vec {k}=0}^{M-n}
\beta ^2_{\vec {m}(\vec {j}-\vec {m}+\vec {k})}\frac {\prod _{i=1}^dC_{j_i+k_i}^{j_i}}{C_M^n}},
\end{eqnarray}
\begin{eqnarray}
|\phi _{\vec {j}}^{[(n+1)...(M+1)]}\rangle 
&=&\sum _{\vec {k}=0}^{M-n}\beta _{\vec {m}(\vec {j}-\vec {m}+\vec {k})}
\sqrt {\frac {\prod _{i=1}^dC_{j_i+k_i}^{j_i}}{C_M^n}}
\nonumber \\
&&|\vec {k}\rangle 
|\vec {j}-\vec {m}+\vec {k}\rangle /\lambda _{\vec {j}}^{[n]}.
\end{eqnarray}
By the same procedure as that of qubit case, we can obtain the following
\begin{eqnarray}
|\phi _{\vec {j}}^{[n...(M+1)]}\rangle 
=\sum _{\alpha _ni_n}\Gamma ^{[n]i_n}_{\vec {j}\alpha _n}
\lambda ^{[n]}_{\alpha _n}|i_n\rangle |\phi _{\alpha _n}^{(n+1)...(M+1)]}\rangle .
\end{eqnarray}
Then we have
\begin{eqnarray}
\Gamma ^{[n]i_n}_{\vec {j}\alpha _n}=\delta _{{\alpha _n}(\vec {j}+\vec {e}_{i_n+1})}
\sqrt {\frac {j_{i_n+1}+1}{n}}/\lambda ^{[n-1]}_{\vec {j}}.
\end{eqnarray}
Still we can define $V^{[n]i_n}_{\alpha _n\alpha _{n-1}}=
\Gamma ^{[n]i_n}_{\alpha _{n-1}\alpha _n}\lambda ^{[n]}_{\alpha _n}$,
and thus we can find that each state $|\Phi _M^{\vec {m}}\rangle $
is a MPS and thus can be sequentially generated. The detailed result
of this part will be presented elsewhere \cite{DF}.

In conclusion, we show that the general $N$ to $M$ universal quantum cloning machine
can be implemented by a sequential manner. Since the sequential generation of 
multipartite state can be implemented in various physical setups such as 
microwave and optical cavity QED, trapped ions and quantum dots etc. This general
sequential quantum cloning machine may be implemented much easier than
the single global implementation scheme. This reduces dramatically the complexity
in implementing the general UQCM. We also show that for $d$-dimensional quantum 
state, the sequential UQCM is also available.    
Besides the universal cloning machine, 
the 1 to $M$ phase-covariant quantum cloning machine can also be 
sequentially implemented. It will be interesting to consider similarly
the general $N$ to $M$ phase-covariant cloning and the economic phase-covariant
cloning. The sequential asymmetric
quantum cloning machine may also be an interesting topic.

{\it Acknowledgements}: HF was supported by "Bairen" program, NSFC and "973" program
(2006CB921107).

{\it Appendix.}--The explicit form of matrices $V$
are presented as:
\begin{eqnarray}
V^{[n]0}_{\alpha _n\alpha _{n-1}}&=&\delta _{\alpha _n\alpha _{n-1}}
\times \nonumber \\
&&\times \left( \frac {\sum _{k=-m}^{M-m-n}X
\frac {C_{M-n}^{m+k}}{C_M^{m+\alpha _{n-1}-1+k}}}
{\sum _{k=-m}^{M-m-n+1}X
\frac {C_{M-n+1}^{m+k}}{C_M^{m+\alpha _{n-1}-1+k}}}
\right) ^{1/2},
\nonumber
\\
V^{[n]0}_{\alpha _n\alpha _{n-1}}&=&\delta _{\alpha _n\alpha _{n-1}+1}
\times \nonumber \\
&&\times \left( \frac {\sum _{k=-m}^{M-m-n}X'
\frac {C_{M-n}^{m+k}}{C_M^{m+\alpha _{n-1}+k}}}
{\sum _{k=-m}^{M-m-n+1}X
\frac {C_{M-n+1}^{m+k}}{C_M^{m+\alpha _{n-1}-1+k}}}
\right) ^{1/2},
\nonumber 
\end{eqnarray}
where notations $X=\beta ^2_{m(\alpha _{n-1}-1+k)}, 
X'=\beta ^2_{m(\alpha _{n-1}+k)}$  are used.
For case $1<n\le M-N+m, \alpha _{n-1}=1,...,n; 
\alpha _n=1,...,(n+1)$, and for case $M-N+m<n\le M-1$,
$\alpha _{n-1}, \alpha _n=1,...,(M-N+m+1)$.  
We can check that
the above defined $V$ satisfies the isometry condition
$\sum _{i_n}\left[ V^{[n]i_n}\right] ^{\dagger }V^{[n]i_n}=1$.
Similarly we have
\begin{eqnarray}
V^{[M]0}_{\alpha _M\alpha _{M-1}}&=&
\delta _{\alpha _M\alpha _{M-1}}
\times \nonumber \\
&&\times \left( \frac {\frac {\beta ^2_{m(\alpha _{M-1}-1-m)}}{C_M^{\alpha _{M-1}}}}
{\frac {\beta ^2_{m(\alpha _{M-1}-1-m)}}{C_{M}^{\alpha _{M-1}-1}}
+\frac {\beta ^2_{m(\alpha _{M-1}-m)}}{C_{M}^{\alpha _{M-1}}}}\right) ^{1/2},
\nonumber
\\
V^{[M]1}_{\alpha _M\alpha _{M-1}}&=&
\delta _{\alpha _M(\alpha _{M-1}+1)}
\times \nonumber \\
&&\times \left( \frac {\frac {\beta ^2_{m(\alpha _{M-1}-m)}}{C_M^{\alpha _{M-1}}}}
{\frac {\beta ^2_{m(\alpha _{M-1}-1-m)}}{C_{M}^{\alpha _{M-1}-1}}
+\frac {\beta ^2_{m(\alpha _{M-1}-m)}}{C_{M}^{\alpha _{M-1}}}}\right) ^{1/2},
\nonumber 
\end{eqnarray} 
where $0\le m\le N-m, \alpha _{M-1}, \alpha _M=1,2,...,(M-N+m+1)$.

For case concerning about ancilla state of the UQCM, assume $1\le l\le M-N$, we have
\begin{eqnarray}
V^{[M+l]0}_{\alpha _{M+l}\alpha _{M+l-1}}
&=&\delta _{\alpha _{M+l}(\alpha _{M+l-1}-1)}\times
\nonumber \\
&&\times \left( \frac {\alpha _{M+l-1}-m-1}{M-N-l+1}\right) ^{1/2},
\nonumber \\
V^{[M+l]1}_{\alpha _{M+l}\alpha _{M+l-1}}
&=&\delta _{\alpha _{M+l}\alpha _{M+l-1}}\times
\nonumber \\
&\times &\left( \frac {M-N-l-\alpha _{M+l-1}+m+1}{M-N-l+1}\right) ^{1/2}.
\nonumber
\end{eqnarray}
$\left( 1\right) $ For $\left( m+1\right) \le \alpha
_{M+l}\le \left( M-N+m-l+1\right) $, $\left( m+2\right) \le
\alpha _{M+l-1}\le \left( M-N+m-l+2\right) $, $V
_{\alpha _{M+l}\alpha _{M+l-1}}^{[M+l]0}=\delta _{\alpha _{M+l}\left( \alpha
_{M+l-1}-1\right) }\sqrt{\frac{\alpha _{M+l-1}-m-1}{M-N-l+1}}$.
For $\alpha _{M+l}=\left( M-N+m-l+2\right) $, $1\le \alpha
_{M+l-1}\leq \left( M-N+m+1\right) $, $V_{\alpha
_{M+l}\alpha _{M+l-1}}^{[M+l]0}=0$.
Otherwise $V_{\alpha _{M+l}\alpha _{M+l-1}}^{[M+l]0}=\delta
_{\alpha _{M+l}\alpha _{M+l-1}}\frac{1}{\sqrt{2}}$.

$\left( 2\right) $ For $\left( m+1\right) \le \alpha _{M+l},\alpha
_{M+l-1}\le \left( M-N+m-l+1\right) $, $V_{\alpha
_{M+l}\alpha _{M+l-1}}^{[M+l]1}=\delta _{\alpha _{M+l}\alpha _{M+l-1}}\sqrt{\frac{%
M-N-l-\alpha _{M+l-1}+m+2}{M-N-l+1}}$.
For $\alpha _{M+l}=\left( M-N+m-l+2\right) $, $1\le \alpha
_{M+l-1}\le \left( M-N+m+1\right) $, $V_{\alpha
_{M+l}\alpha _{M+l-1}}^{[M+l]0}=0$.
Otherwise $V_{\alpha _{M+l}\alpha _{M+l-1}}^{[M+l]0}=\delta
_{\alpha _{M+l}\alpha _{M+l-1}}\frac{1}{\sqrt{2}}$.

\end{document}